\documentclass[a4paper,12pt]{article}
\usepackage{amsmath,amssymb,amsfonts,color,comment}
\usepackage[normalem]{ulem}
\usepackage{cancel}
\usepackage{slashed, tensor, bm, physics}
\usepackage{caption}
\usepackage{graphicx} 
\usepackage{multirow}
\usepackage{float}
\usepackage{jheppub} 
\usepackage{xcolor}

\usepackage{booktabs} 

\title{Kaon Portal to Freeze-in Dark Matter}

\author[a,b]{Motoi Endo,}
\author[c]{Takumu Yamanaka}

\affiliation[a\,]{KEK Theory Center, IPNS, KEK, Tsukuba 305--0801, Japan}
\affiliation[b\,]{Graduate Institute for Advanced Studies, SOKENDAI, Tsukuba, Ibaraki 305--0801, Japan}
\affiliation[c\,]{Department of Physics, The University of Osaka, Machikaneyama-cho, Toyonaka, Osaka 560-0043, Japan}

\abstract{
We investigate freeze-in production of light dark matter through the quark flavor-changing operator $(\bar{s}\gamma_\mu d)(\bar{\chi}\gamma^\mu\chi)$ in a low-reheating cosmology.
For reheating temperatures below the QCD crossover, kaon decays and scatterings generate the dark matter abundance through $K\to\pi\chi\bar{\chi}$ and $K\pi\to\chi\bar{\chi}$.
The same interaction induces the rare kaon decays $K^+\to\pi^+\chi\bar{\chi}$ and $K_L\to\pi^0\chi\bar{\chi}$.
This links the freeze-in relic abundance to searches at NA62, KOTO, and KOTO II.
We find that lower reheating temperatures require larger couplings to compensate for the Boltzmann-suppressed kaon abundance, making kaon-driven freeze-in dark matter testable at rare kaon decay experiments.
}

\begin{document}
\begin{flushright}
KEK--TH--2831\\
OU--HET--1313
\end{flushright}
\maketitle

\section{Introduction}
\label{sec:introduction}

The origin of dark matter (DM) remains one of the central open questions in particle physics and cosmology.
Although the paradigm of weakly-interacting massive particles provides a simple and predictive framework, the absence of convincing signals, in particular from direct detection experiments~\cite{XENON:2025vwd, LZ:2024zvo, PandaX:2024qfu}, has motivated growing interest in alternative scenarios.

A compelling possibility is the freeze-in mechanism~\cite{McDonald:2001vt, Hall:2009bx}, in which the DM particle never reaches thermal equilibrium with the Standard Model (SM) bath.
Instead, it is gradually populated through decays and annihilations of SM particles, with the yield determined by an extremely small coupling to the visible sector.
This feebleness, however, poses a fundamental observational challenge. 
If the coupling is too small to be probed in laboratory experiments, the freeze-in scenario becomes difficult to test directly.

A crucial observation is that the coupling required to reproduce the observed relic abundance depends on the reheating temperature $T_{\rm RH}$.
When $T_{\rm RH}$ is low compared with the relevant mass scales of the DM production processes, the Boltzmann suppression of the initial-state particles requires a larger coupling to reproduce the observed relic abundance, potentially bringing it within experimental reach (cf.~Refs.~\cite{Bhattiprolu:2022sdd, Boddy:2024vgt, Bernal:2024ndy})\footnote{Similar enhancement of couplings due to low reheating temperatures has been explored in the context of leptophilic effective operators and lepton collider phenomenology~\cite{Barman:2024nhr, Barman:2024tjt, Borah:2025ema}.}.
This motivates considering cosmological histories with low reheating temperatures and identifying the SM degrees of freedom that dominate DM production during this epoch.

In this paper, we focus on a DM candidate $\chi$ that couples to the SM through the quark flavor-changing  operator,
\begin{equation}
  \mathcal{L}_{\text{int}} = \frac{C_V}{\Lambda^2}(\bar{s}\gamma^\mu d)(\bar{\chi}\gamma_\mu \chi) + \text{h.c.},
  \label{eq:operator}
\end{equation}
where $\chi$ is a Dirac fermion with mass $m_\chi$, and $C_V/\Lambda^2$ parametrizes the coupling strength.
When $T_{\rm RH}$ falls below the QCD crossover temperature $T_{\rm QCD} \sim 150$--$160\,{\rm MeV}$~\cite{Aoki:2006br, HotQCD:2018pds}, DM production through this interaction takes place in the hadronic bath.\footnote{\label{footnote:model}
Such low reheating temperatures can arise in well-motivated extensions of the SM. 
Examples include long-lived moduli fields in supergravity or string compactifications~\cite{Coughlan:1983ci, deCarlos:1993wie, Acharya:2008bk, Allahverdi:2013noa}, and flatons in thermal inflation~\cite{Lyth:1995ka}. 
These models generally involve a non-instantaneous reheating era, during which DM can also be produced. In this paper we assume instantaneous reheating (see footnote~\ref{footnote:TRH}).
On the other hand, the observed baryon asymmetry can be generated, for example, via Affleck-Dine baryogenesis, possibly accompanied by Q-ball formation and decay~\cite{Affleck:1984fy, Dine:1995uk, Coleman:1985ki, Kusenko:1997si, Enqvist:1997si}.
}
Kaons then generate DM through the decays $K \to \pi\chi\bar{\chi}$ and the scatterings $K\pi \to \chi\bar{\chi}$.
The same operator induces the rare kaon decays $K^+ \to \pi^+\chi\bar{\chi}$ and $K_L \to \pi^0\chi\bar{\chi}$, which can be searched for at NA62~\cite{NA62:2024pjp, Chang:2026vvx} and KOTO~\cite{KOTO:2024zbl}.
The relic density and the rare decay branching ratios are controlled by the same interaction.
This provides a direct connection between the low-reheating cosmology and laboratory searches.
We solve the Boltzmann equation for the DM yield including the leading kaon decay and scattering contributions and identify the parameter regions where the freeze-in abundance reproduces the observed relic density $\Omega h^2 = 0.12$~\cite{Planck:2018vyg}.
We show that the freeze-in bottleneck, {\it i.e.,} the coupling feebleness required in conventional freeze-in scenarios, can be alleviated in this scenario.

The experimental testability of freeze-in DM has been discussed in several contexts.
In models with light mediators, direct detection experiments can probe the feeble interactions responsible for freeze-in production~\cite{Hambye:2018dpi}.
Reference \cite{Bhattiprolu:2022sdd} studied hadrophilic UV freeze-in at low reheating temperatures, focusing on pion-bath production and direct detection signatures.
Freeze-in has also been connected to flavor physics in flavon and Froggatt-Nielsen portal models~\cite{Babu:2023zni, Mandal:2023jnv}, and in lepton-flavor-violating scenarios involving fermions~\cite{DAmbrosio:2021wpd} or axion-like particles~\cite{Panci:2022wlc}.
However, these studies do not directly connect quark flavor-changing interactions to both the freeze-in relic abundance and rare meson observables.

Rare meson decays with missing energy have been analyzed as probes of light dark-sector particles coupled through flavor-changing neutral-current interactions. 
Such probes have been studied systematically in $K$ and $B$ decays~\cite{Kamenik:2011vy}, including dark-sector fermion pairs in rare kaon decays~\cite{Fabbrichesi:2019bmo} and light invisible particles in $B$ and $K$ meson decays~\cite{He:2022ljo}.
These works establish rare meson decays as powerful probes of light invisible states, though they do not address the cosmological production of the DM abundance through the same kaon interactions.
Our work bridges this gap.  
For $T_{\rm RH} < T_{\rm QCD}$, the freeze-in relic density is generated by kaon processes, $K \to \pi\chi\bar{\chi}$ and $K\pi \to \chi\bar{\chi}$, while the same operator governs the rare decays $K^+ \to \pi^+\chi\bar{\chi}$ and $K_L \to \pi^0\chi\bar{\chi}$ searched for at NA62 and KOTO.

Before proceeding, we comment on the cosmological lower bound on the reheating temperature. 
Very low reheating scenarios are constrained by BBN and CMB/LSS observables, mainly through incomplete neutrino thermalization and the resulting changes in light-element abundances and $N_{\rm eff}$. 
Recent analyses require $T_{\rm RH}$ to be above a few MeV~\cite{Kawasaki:2000en, Hannestad:2004px, deSalas:2015glj, Hasegawa:2019jsa, Barbieri:2025moq}, with the strongest current bound being $T_{\rm RH} \gtrsim 6\,{\rm MeV}$ for the standard electromagnetic reheating scenario. 
This lower limit is well below the reheating temperatures used in our analysis.
We restrict ourselves to $T_{\rm RH} < T_{\rm QCD}$, where the relevant degrees of freedom for freeze-in production are hadrons.

The remainder of this paper is organized as follows.
In section \ref{sec:Kaon_decay}, we derive the decay rates for the rare kaon processes induced by the quark flavor-changing interaction with DM.
In section \ref{sec:freeze-in}, we formulate the Boltzmann equation for freeze-in production from kaon decays and scatterings.
In section \ref{sec:result}, we present the numerical results and compare the viable freeze-in parameter space with current and future rare-kaon searches.
Section \ref{sec:conclusion} is devoted to conclusions.

\section{Kaon decay}
\label{sec:Kaon_decay}

In this section, we derive the decay rates of kaons into a pion and a pair of DM particles induced by the operator in Eq.~\eqref{eq:operator}. 
These decays determine the laboratory signals in rare kaon decay experiments and contribute to the freeze-in abundance. 
We work at leading order in isospin symmetry and neglect the small CP violation in neutral-kaon mixing unless otherwise stated.

\subsection{Charged Kaon}

From Eq.~\eqref{eq:operator}, the transition amplitude for $K^+\to\pi^+\chi\bar\chi$ is given by
\begin{align}
    i{\cal M}(K^+\to\pi^+\chi\bar\chi) &= 
    \frac{iC_V}{\Lambda^2}
    \matrixel{\pi^+(p_\pi)}{\bar s\gamma^\mu d}{K^+(p_K)}
    \bar u_\chi(p_1)\gamma_\mu v_\chi(p_2),
    \label{eq:amplitude}
\end{align}
where $p_K, p_\pi, p_1$, and $p_2$ are the four-momenta of $K$, $\pi$, $\chi$, and $\bar\chi$, respectively.
The hadronic matrix element can be decomposed in terms of form factors as
\begin{align}
    \matrixel{\pi^+(p_\pi)}{\bar s\gamma^\mu d}{K^+(p_K)} = 
    f_+(q^2)(p_K+p_\pi)^\mu+f_-(q^2)q^\mu,
    \label{eq:formfactor}
\end{align}
where $q=p_K-p_\pi=p_1+p_2$. 
We parameterize the vector form factor as
\begin{align}
 f_+(q^2) &= f_+(0)\left[1+\lambda\frac{q^2}{m_\pi^2}\right],
 \label{eq:formfactor2}
\end{align}
with $f_+(0) = 0.97$ and $\lambda = 0.025$~\cite{FlavourLatticeAveragingGroupFLAG:2024oxs, Boito:2008fq, Boito:2010me}.
This linear parametrization is appropriate for the region $0 \leq q^2 \leq (m_K - m_\pi)^2 \simeq 0.13~{\rm GeV}^2$ relevant for the kaon decay.
The contribution proportional to $f_-(q^2)$ in Eq.~\eqref{eq:amplitude} vanishes, since $q^\mu\bar u_\chi(p_1)\gamma_\mu v_\chi(p_2)=0$.

The differential decay rate is therefore obtained as
\begin{align}
    \dv{\Gamma(K^+\to\pi^+\chi\bar\chi)}{q^2} = 
    \frac{|C_V|^2}{192\pi^3m_K^3\Lambda^4} 
    |f_+(q^2)|^2 \, \beta_\chi(q^2) \, \lambda_K^{\frac{3}{2}}(q^2)
    \left(1+\frac{2m_\chi^2}{q^2}\right),
    \label{eq:diffdec_charge}
\end{align}
where 
\begin{align}
    \beta_\chi(q^2) = \sqrt{1-\frac{4m_\chi^2}{q^2}}, ~~~
    \lambda_K(q^2) = \lambda(m_K^2,m_\pi^2,q^2),
\end{align}
and $\lambda(a,b,c)=a^2+b^2+c^2-2ab-2bc-2ca$.
The total decay rate becomes
\begin{align}
    \Gamma = \int \dd q^2\dv{\Gamma}{q^2}.
    \label{eq:chargedKaonDecay}
\end{align}

The latest preliminary result from NA62, combining the data collected in 2016--2024, is
\begin{align}
    {\rm BR}(K^+\to\pi^+\nu\bar\nu)_{\rm NA62}
    = \left(9.6^{+1.9}_{-1.8}\right)\times 10^{-11},
    \label{eq:na62_latest}
\end{align}
with a precision better than $20\%$~\cite{Chang:2026vvx}.
Since the signal topology of $K^+\to\pi^+\chi\bar\chi$ is the same one-track plus missing-energy final state as $K^+\to\pi^+\nu\bar\nu$, this measurement provides a sensitive probe of the charged-kaon channel.
In applying the NA62 result to the DM mode, however, one should note that the experimental acceptance is not strictly identical to that for the SM neutrino mode~\cite{Fabbrichesi:2019bmo}.
For light $\chi$, the missing-mass distribution overlaps the NA62 signal regions and the acceptance is expected to be comparable.
For larger $m_\chi$, the spectrum starts at $q^2=4m_\chi^2$ and can be shifted toward different regions of pion momentum and missing mass.
Thus, a precise bound requires a dedicated recast with the NA62 efficiency maps.
In the numerical analysis below, we use Eq.~\eqref{eq:na62_latest} as a proxy for the current NA62 sensitivity to the charged-kaon mode and do not include additional acceptance corrections.

\subsection{Neutral Kaon}

In this subsection, we derive the decay rates for neutral kaons.
For the transition $K_L\to\pi^0\chi\bar\chi$, we neglect the small CP violation in neutral-kaon mixing,
\begin{align}
    \ket{K_L} \simeq \frac{1}{\sqrt{2}} \bigl(\ket{K^0}+\ket{\bar K^0}\bigr),
\end{align}
where we use the convention $(CP) | K^0 \rangle = -| \bar K^0 \rangle$ under the CP transformation.
The decay amplitude is then written as
\begin{align}
    i{\cal M}(K_L\to\pi^0\chi\bar\chi)
    &=
    \frac{i}{\Lambda^2}
    \matrixel{\pi^0(p_\pi)}{(C_V\bar s\gamma^\mu d+C_V^*\bar d\gamma^\mu s)}{K_L(p_K)}
    \bar u_\chi(p_1)\gamma_\mu v_\chi(p_2) \notag\\
    &= \frac{\sqrt{2}i\Im C_V}{\Lambda^2}
    \matrixel{\pi^0(p_\pi)}{\bar s\gamma^\mu d}{K^0(p_K)}
    \bar u_\chi(p_1)\gamma_\mu v_\chi(p_2),
\end{align}
where we use $\matrixel{\pi^0}{\bar s\gamma^\mu d}{K^0} = - \matrixel{\pi^0}{\bar d\gamma^\mu s}{\bar K^0}$.
Using the isospin relation, the hadronic matrix element is related to that in Eq.~\eqref{eq:formfactor} as
\begin{align}
    \matrixel{\pi^0(p_\pi)}{\bar s\gamma^\mu d}{K^0(p_K)} = -
    \frac{1}{\sqrt{2}} \Bigl[ f_+(q^2)(p_K+p_\pi)^\mu+f_-(q^2)q^\mu \Bigr].\label{eq:neutral_matrix_element}
\end{align}
As in the charged kaon decay, the contribution proportional to $f_-(q^2)$ vanishes. 
We therefore obtain
\begin{align}
    \dv{\Gamma(K_L\to\pi^0\chi\bar\chi)}{q^2} &=
    \frac{(\Im C_V)^2}{192\pi^3m_K^3\Lambda^4}
    |f_+(q^2)|^2 \, \beta_\chi(q^2) \, \lambda_K^{\frac{3}{2}}(q^2)
    \left(1+\frac{2m_\chi^2}{q^2}\right).
\end{align}

The current best bound on the neutral mode is set by KOTO using the data taken in 2021~\cite{KOTO:2024zbl},
\begin{align}
    {\rm BR}(K_L\to\pi^0\nu\bar\nu)_{\rm KOTO}
    < 2.2\times 10^{-9}
    \quad (90\%~{\rm C.L.}) .
    \label{eq:koto_current}
\end{align}
This search observed no events in the signal region and improved the previous KOTO limit.
With further data accumulation, the KOTO program is expected to reach sensitivities below $10^{-10}$, while the proposed KOTO II experiment aims at a sensitivity around $10^{-12}$ or below~\cite{KOTO:2025gvq}.

The same acceptance caveat as in the charged kaon mode applies to the neutral kaon mode.
For light $\chi$, the signal acceptance is expected to be close to that for $K_L\to\pi^0\nu\bar\nu$, while it can change for larger $m_\chi$ because the reconstructed $\pi^0$ kinematics are shifted.
In the numerical analysis, we use Eq.~\eqref{eq:koto_current} as a proxy for the present KOTO sensitivity and do not include acceptance corrections.

\section{Freeze-in DM}
\label{sec:freeze-in}

In this section, we formulate the freeze-in production of $\chi$ in a low-reheating cosmology.
We consider temperatures below the QCD crossover and well above the BBN lower bound on the reheating temperature.
The relevant degrees of freedom are therefore hadrons, rather than quarks and gluons~\cite{Bhattiprolu:2022sdd}.
At $T\sim 100\,{\rm MeV}$, elastic scatterings with ambient pions occur much more rapidly than $K^0$--$\bar K^0$ oscillations.
The neutral kaons in the thermal bath should therefore be treated as the flavor eigenstates $K^0$ and $\bar K^0$, rather than as coherently propagating $K_L$ and $K_S$ states~\cite{Amelino-Camelia:1999uhg}.
The DM abundance is then generated through the kaon processes
\begin{align}
    & K^+ \to \pi^+\chi\bar\chi, ~~~
    K^0 \to \pi^0\chi\bar\chi, \notag\\
    & K^+\pi^- \to \chi\bar\chi, ~~~
    K^0\pi^0 \to \chi\bar\chi,
    \label{eq:process}
\end{align}
together with their charge-conjugate processes.\footnote{
Heavier $s$-quark hadrons are Boltzmann suppressed relative to kaons for $T_{\rm RH}=60$--$100\,{\rm MeV}$. 
The $K^*(892)$ contribution is included through the timelike $K\pi$ vector form factor in the scattering process, and adding a thermal $K^*$ initial state separately would double count the resonant contribution.
}
We assume that the mesons appearing in the initial states follow their equilibrium thermal distributions in the hadronic bath through strong and electromagnetic interactions. 
Their number densities are therefore well described by Maxwell--Boltzmann distributions, with kaons and pions Boltzmann suppressed for $T<m_K$ and $T<m_\pi$, respectively. 
Since the interaction responsible for DM production is feeble, we neglect inverse processes and quantum-statistical factors in the collision terms.

\subsection{DM Production}

We first consider DM production from kaon decays.
Among the processes in Eq.~\eqref{eq:process}, the charged kaon decay rate is given in Eq.~\eqref{eq:diffdec_charge}.
For the neutral kaon decay $K^0\to\pi^0\chi\bar\chi$, the isospin relation gives the differential decay rate for $K^0$ as
\begin{align}
    \dv{\Gamma(K^0\to\pi^0\chi\bar\chi)}{q^2} = 
    \frac{1}{2}\dv{\Gamma(K^+\to\pi^+\chi\bar\chi)}{q^2}.
    \label{eq:diffdec_neutral}
\end{align}

Next, we consider DM production from kaon scattering processes.
The scattering amplitude for $K^+\pi^-\to\chi\bar\chi$ is 
\begin{align}
    i{\cal M}(K^+\pi^-\to\chi\bar\chi) &= 
    \frac{iC_V}{\Lambda^2}
    \matrixel{0}{\bar s \gamma^\mu d}{K^+(p_K)\pi^-(p_\pi)}\bar u_\chi(p_1)\gamma_\mu v_\chi(p_2).
\end{align}
The hadronic matrix element is related by crossing symmetry to the $K\to\pi$ form factors in Eq.~\eqref{eq:formfactor}. 
We therefore parameterize it as
\begin{align}
    \matrixel{0}{\bar s \gamma^\mu d}{K^+(p_K)\pi^-(p_\pi)}
    = F_+(s)(p_K-p_\pi)^\mu +F_-(s)(p_K+p_\pi)^\mu.
\end{align}
The contribution proportional to $F_-(s)$ vanishes after contraction with the DM vector current, since $(p_1+p_2)^\mu\bar u_\chi(p_1)\gamma_\mu v_\chi(p_2)=0$.
Unlike the form factors for the kaon decay processes discussed in Section~\ref{sec:Kaon_decay}, the scattering process $K\pi\to\chi\bar\chi$ probes the timelike region starting at $s=(m_K+m_\pi)^2\simeq 0.40\,{\rm GeV}^2$.
In the collision term of the Boltzmann equation, large values of $\sqrt{s}$ are exponentially suppressed by the thermal factor $K_1(\sqrt{s}/T)$.
Since we consider reheating temperatures below the QCD crossover, the dominant contribution comes from the low-energy region near the $K\pi$ threshold and the $K^*(892)$ resonance.
We therefore use a Breit-Wigner form as a simple one-resonance approximation to the $K\pi$ vector form factor~\cite{Jamin:2006tk, Boito:2008fq, Boito:2010me},
\begin{align}
    F_+(s) &\simeq 
    f_+(0)\frac{m_{K^*}^2}{m_{K^*}^2-s-im_{K^*}\gamma_{K^*}(s)},
    \label{eq:form_factor_scat} \\
    \gamma_{K^*}(s) &= 
    \gamma_{K^*}\frac{m_{K^*}}{\sqrt{s}} \left[ \frac{q_{K\pi}(s)}{q_{K\pi}(m_{K^*}^2)} \right]^3,
    \label{eq:form_factor_gamma} \\
    q_{K\pi}(s) &= \frac{1}{2\sqrt{s}} \lambda^{\frac{1}{2}}(s,m_K^2,m_\pi^2),
\end{align}
where $m_{K^*} = 892.0\,{\rm MeV}$ and $\gamma_{K^*} = 46.5\,{\rm MeV}$ denote the pole mass and width of the $K^*(892)$ resonance, as extracted from the dispersive analyses of Refs.~\cite{Boito:2008fq, Boito:2010me}.
The Breit-Wigner form is normalized such that $F_+(0)=f_+(0)$ in the zero-width, subthreshold extrapolation.
The energy-dependent width in Eq.~\eqref{eq:form_factor_gamma} is used only in the physical $K\pi$ region, $s\ge (m_K+m_\pi)^2$, relevant for the scattering collision term.
Below the threshold, the real $K\pi$ channel is closed, and the width is set to zero in this simple parametrization.

The scattering cross section is therefore obtained as
\begin{align}
    \sigma(K^+\pi^-\to\chi\bar\chi)
    &= \frac{|C_V|^2|F_+(s)|^2}{12\pi\Lambda^4}
    \lambda^{\frac{1}{2}}(s,m_K^2,m_\pi^2)
    \sqrt{1-\frac{4m_\chi^2}{s}}
    \left(1+\frac{2m_\chi^2}{s}\right).
    \label{eq:crosssec_charge}
\end{align}
Using the isospin relation, the cross section for $K^0\pi^0\to\chi\bar\chi$ is given by 
\begin{align}
    \sigma(K^0\pi^0\to\chi\bar\chi)&=\frac{1}{2}\sigma(K^+\pi^-\to\chi\bar\chi).
    \label{eq:crosssec_neutral}
\end{align}

\subsection{Boltzmann Equation}

We now combine the decay rates and scattering cross sections derived above into the Boltzmann equation for the DM yield.
Since the interaction is feeble, the DM abundance remains far below thermal equilibrium and inverse processes can be neglected.
The evolution of the yield is then governed by
\begin{align}
    \dv{Y_\chi}{x}&=\frac{2}{\tilde sHx}\left(1-\frac{x}{3g_{*s}}\dv{g_{*s}}{x}\right)[{\cal C}_{\rm scat}^{K^+\pi^-}(T)+{\cal C}_{\rm dec}^{K^+}(T)+{\cal C}_{\rm scat}^{K^0\pi^0}(T)+{\cal C}_{\rm dec}^{K^0}(T)],\label{eq:boltzmann}
\end{align}
where $x=m_K/T$, and $Y_\chi=n_\chi/\tilde s$ denotes the yield of $\chi$.
The same Boltzmann equation holds for $Y_{\bar\chi}$.
The overall factor of 2 on the right-hand side accounts for the charge-conjugate kaon processes.
Here, $g_{*s}(T)$ and $g_{*}(T)$ denote the effective numbers of relativistic degrees of freedom for entropy and energy densities, respectively.
In the numerical analysis, we include the temperature dependence of $g_*(T)$ and $g_{*s}(T)$~\cite{Saikawa:2020swg}.
We assume radiation domination during freeze-in production and use
\begin{align}
    H(T) =
    \sqrt{\frac{\pi^2 g_*(T)}{90}}\,\frac{T^2}{M_{\rm Pl}},~~~
    \tilde s(T) =
    \frac{2\pi^2}{45}g_{*s}(T)T^3,
\end{align}
where $M_{\rm Pl}$ is the reduced Planck mass.

For the scattering processes, the cross section entering the collision term is weighted by the thermal distributions of the initial-state mesons.
Using the Maxwell-Boltzmann approximation for the kaon and pion distributions, the result can be written as a one-dimensional integral over the invariant mass $s$ of the $K\pi$ system~\cite{Gondolo:1990dk, Edsjo:1997bg},
\begin{align}
    {\cal C}_{\rm scat}
    &=\frac{T}{32\pi^4}\int\dd s\,\sigma(s)\frac{\lambda(s,m_K^2,m_\pi^2)}{\sqrt{s}}K_1\left(\frac{\sqrt{s}}{T}\right),
    \label{eq:colscat}
\end{align}
where $\sigma(s)$ denotes the scattering cross section in Eqs.~\eqref{eq:crosssec_charge} and~\eqref{eq:crosssec_neutral}, and $K_1(z)$ is the modified Bessel function of the second kind.

For the decay processes, the collision term is obtained by thermally averaging the decay rate in Eqs.~\eqref{eq:diffdec_charge} and \eqref{eq:diffdec_neutral}.
In the Maxwell-Boltzmann approximation, this gives
\begin{align}
    {\cal C}_{\rm dec}
    &=\frac{m_K^2T}{2\pi^2}K_1\left(\frac{m_K}{T}\right)\int \dd q^2\dv{\Gamma}{q^2}.
    \label{eq:coldecay}
\end{align}

Consequently, the collision terms entering Eq.~\eqref{eq:boltzmann} are expressed as
\begin{align}
    {\cal C}_{\rm scat}^{K^+\pi^-}&=\frac{T}{32\pi^4}\int_{(m_K+m_\pi)^2}^\infty \dd s\, \sigma(K^+\pi^-\to\chi\bar\chi)\frac{\lambda(s,m_K^2,m_\pi^2)}{\sqrt{s}}K_1\left(\frac{\sqrt{s}}{T}\right),\\
    {\cal C}_{\rm scat}^{K^0\pi^0}&=\frac{T}{32\pi^4}\int_{(m_K+m_\pi)^2}^\infty \dd s\, \sigma(K^0\pi^0\to\chi\bar\chi)\frac{\lambda(s,m_K^2,m_\pi^2)}{\sqrt{s}}K_1\left(\frac{\sqrt{s}}{T}\right),\\
    {\cal C}_{\rm dec}^{K^+}&=\frac{m_K^2T}{2\pi^2}K_1\left(\frac{m_K}{T}\right)\int_{4m_\chi^2}^{(m_K-m_\pi)^2}\dd q^2\dv{\Gamma(K^+\to\pi^+\chi\bar\chi)}{q^2},\\
    {\cal C}_{\rm dec}^{K^0}&=\frac{m_K^2T}{2\pi^2}K_1\left(\frac{m_K}{T}\right)\int_{4m_\chi^2}^{(m_K-m_\pi)^2} \dd q^2\dv{\Gamma(K^0\to\pi^0\chi\bar\chi)}{q^2}.
\end{align}

We solve Eq.~\eqref{eq:boltzmann} with the initial condition $Y_\chi(x_{\rm RH})=0$, where $x_{\rm RH}=m_K/T_{\rm RH}$.
We define $T_{\rm RH}$ as the temperature at the onset of radiation domination and do not include possible DM production before this epoch. 
This prescription corresponds to the assumption of instantaneous reheating, in which the maximum temperature of the radiation bath is identical to $T_{\rm RH}$.\footnote{\label{footnote:TRH}
Such production depends on the UV completion and on the maximum temperature before radiation domination. 
In a non-instantaneous reheating era, such as an early matter-dominated phase driven by the long-lived moduli fields or flatons mentioned in footnote~\ref{footnote:model}, DM is produced before the completion of reheating~\cite{Co:2015pka,Drees:2017iod,Banerjee:2022fiw}. 
In such cases, the predicted coupling that reproduces the DM relic abundance becomes smaller than our numerical results shown in section~\ref{sec:result}. 
}
Hence, our calculation gives the freeze-in abundance generated after reheating has completed.
Since both $\chi$ and $\bar\chi$ contribute to the DM energy density, the relic abundance is given by
\begin{align}
    \Omega h^2&=\frac{2m_\chi \tilde s_0 Y_\chi(x=\infty)}{\rho_{c,0}}\simeq5.49\times 10^8~{\rm GeV^{-1}}\times m_\chi Y_\chi(x=\infty),
\end{align}
where $\rho_{c,0}\simeq 1.053\times 10^{-5}h^2\,{\rm GeV\, cm^{-3}}$ and $\tilde s_0=2891.2\,{\rm cm^{-3}}$ are the critical energy density and the entropy density in the current universe, respectively~\cite{ParticleDataGroup:2024cfk}.
The overall factor of 2 on the right-hand side accounts for the energy density carried by both $\chi$ and $\bar\chi$.

\section{Numerical Results}
\label{sec:result}

We evaluate the freeze-in abundance and the rare kaon decay rates.
We solve the Boltzmann equation in Eq.~\eqref{eq:boltzmann} from the initial temperature $T_{\rm RH}$ down to sufficiently low temperatures at which the kaon abundance becomes negligible.
For $K_L \to \pi^0\chi\bar\chi$, we take $(\mathrm{Im}\,C_V)^2/|C_V|^2=1/2$ as a representative choice, corresponding to a CP-violating phase of order unity.

\begin{figure}[tbp]
    \centering
    \includegraphics[width=0.8\linewidth]{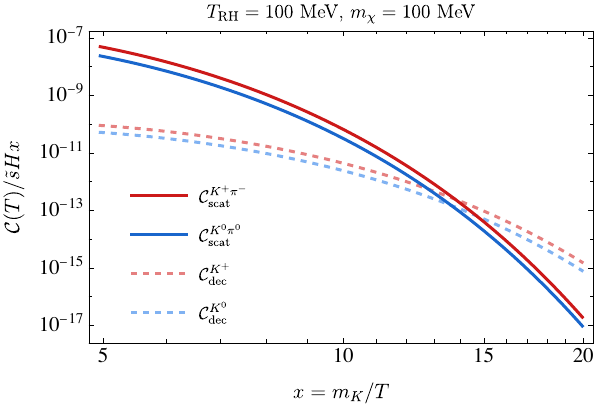}
    \caption{Individual contributions to normalized collision terms as a function of $x=m_K/T$, for $T_{\rm RH}=100\,{\rm MeV}$ and $m_\chi=100\,{\rm MeV}$. 
    The solid red and blue curves show the contributions of $K^+\pi^-\to\chi\bar\chi$ and $K^0\pi^0\to\chi\bar\chi$, respectively, while the dashed red and blue curves show the contributions of $K^+\to\pi^+\chi\bar\chi$ and $K^0\to\pi^0\chi\bar\chi$. 
    In this plot, we fix $\Lambda=100\,{\rm TeV}$ and $|C_V|=\sqrt{2}$.}
    \label{fig:collision}
\end{figure}

Figure~\ref{fig:collision} compares the four contributions to the collision term, normalized as ${\cal C}/(\tilde s H x)$.
At high temperature, the scattering contributions dominate because the initial-state pion is still thermally populated and the $K \pi$ cross section benefits from the timelike $K \pi$ form factor, in particular by the nearby $K^*(892)$ resonance, whereas the three-body kaon decay is limited by the available phase space.
At lower temperatures, the scattering rates decrease more rapidly because the pion in the initial state gives an additional Boltzmann suppression.
The decay contributions then become relatively more important.
The charged channels are larger than the corresponding neutral channels, as expected from the isospin factors in Eqs.~\eqref{eq:diffdec_neutral} and \eqref{eq:crosssec_neutral}.

\begin{figure}[tbp]
    \centering
    \includegraphics[width=\linewidth]{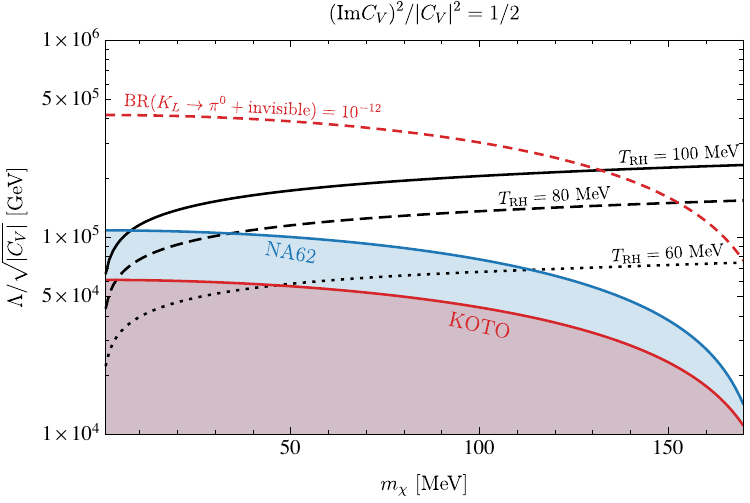}
    \caption{Freeze-in relic abundance of DM and branching ratios of kaon decays in the $(m_\chi,\,\Lambda/\sqrt{|C_V|})$ plane.
    We fix $(\mathrm{Im}\,C_V)^2/|C_V|^2=1/2$.
    The black dotted, dashed, and solid curves show the contours along which the freeze-in mechanism reproduces the observed relic abundance, $\Omega h^2=0.12$, for reheating temperatures $T_{\rm RH}=60,\,80,\,100~{\rm MeV}$, respectively.
    The blue and red shaded regions show proxy sensitivities derived from the current NA62 measurement of ${\rm BR}(K^+\to\pi^+\nu\bar\nu)$~\cite{NA62:2024pjp,Chang:2026vvx} and the current KOTO bound on ${\rm BR}(K_L\to\pi^0\nu\bar\nu)$~\cite{KOTO:2024zbl}, respectively. 
    The red dashed curve indicates the projected KOTO II sensitivity, for which we assume ${\rm BR}(K_L\to\pi^0+\mathrm{invisible})=10^{-12}$~\cite{KOTO:2025gvq}.
    }
    \label{fig:lambda_vs_mchi}
\end{figure}

Figure~\ref{fig:lambda_vs_mchi} shows the freeze-in target and the rare kaon decay sensitivities in the $(m_\chi,\Lambda/\sqrt{|C_V|})$ plane.
The black dotted, dashed, and solid curves correspond to $\Omega h^2=0.12$ for $T_{\rm RH}=60$, $80$, and $100\,{\rm MeV}$, respectively.
Lower reheating temperatures require larger interaction strengths because the kaon abundance is more strongly Boltzmann suppressed.
The freeze-in target therefore moves toward smaller values of $\Lambda/\sqrt{|C_V|}$.
The current NA62 and KOTO results already reach part of the target region, especially for $T_{\rm RH}=60\,{\rm MeV}$.
The projected KOTO II sensitivity can probe a much larger part of the freeze-in target for $T_{\rm RH}=80$ and $100\,{\rm MeV}$.
This shows that rare kaon decay experiments can test kaon-driven freeze-in scenarios.

Although a dedicated analysis of small-scale structure would require the non-thermal momentum distribution of $\chi$, studies of freeze-in DM typically find lower mass bounds in the keV range from Lyman-$\alpha$ forest and related structure-formation constraints~\cite{DEramo:2020gpr,Decant:2021mhj,DEramo:2025jsb}.
We therefore consider the conservative mass range $1\,{\rm MeV}\leq m_\chi\leq170\,{\rm MeV}$.
We have checked that the interaction rate of $\chi$, estimated from the collision term in the Boltzmann equation, remains below the Hubble expansion rate throughout the parameter region shown.
This confirms that $\chi$ is never thermalized with the SM bath in this region.

\begin{figure}[tbp]
\centering
        \includegraphics[width=0.75\linewidth]{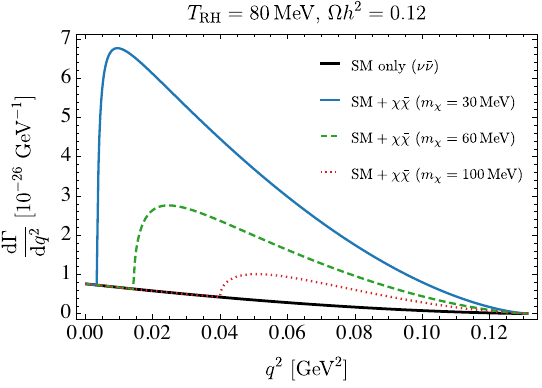}
        \\[5mm]
        \includegraphics[width=0.75\linewidth]{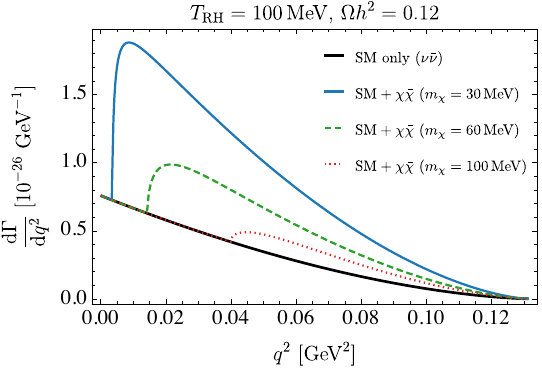}
    \caption{Differential decay rate for $K_L \to \pi^0+\mathrm{invisible}$ as a function of $q^2$.
    The black solid curve shows the SM contribution from $K_L\to\pi^0\nu\bar\nu$~\cite{Buras:2015qea}, while the colored curves show the sum of the SM contribution and the DM contribution from $K_L\to\pi^0\chi\bar\chi$ for $m_\chi = 30, 60$, and $100\,{\rm MeV}$.
    The top and bottom panels correspond to $T_{\rm RH}=80\,{\rm MeV}$ and $100\,{\rm MeV}$, respectively.
    For each value of $m_\chi$ and $T_{\rm RH}$, $\Lambda/\sqrt{|C_V|}$ is fixed by requiring $\Omega h^2 = 0.12$.}
    \label{fig:dGamma_q2}
\end{figure}

The invisible invariant-mass spectrum provides information beyond the inclusive branching ratio.
In the neutral mode, this spectrum is reflected in the reconstructed $\pi^0$ kinematics, although a dedicated experimental analysis is required to map it onto the KOTO acceptance.
Figure~\ref{fig:dGamma_q2} shows the differential spectrum for $K_L\to\pi^0+{\rm invisible}$, with $\Lambda/\sqrt{|C_V|}$ fixed by the relic density condition.
The DM contribution turns on at $q^2=4m_\chi^2$.
Larger DM masses shift the excess toward larger missing invariant masses.
For $T_{\rm RH}=80\,{\rm MeV}$, the relic-density condition requires a larger coupling and produces a visible distortion of the spectrum.
For $T_{\rm RH}=100\,{\rm MeV}$, the required coupling is smaller and the deviation from the SM spectrum is reduced.

\section{Conclusions}
\label{sec:conclusion}

We have studied freeze-in production of light DM through the quark flavor-changing operator
$(\bar{s}\gamma_\mu d)(\bar{\chi}\gamma^\mu\chi)$ in a low-reheating cosmology.
For reheating temperatures below the QCD crossover, the relevant SM degrees of freedom are hadrons rather than quarks and gluons, and the DM abundance is generated by kaon decays and scattering processes.
We derived the decay rates for $K\to\pi\chi\bar\chi$ and the cross sections for
$K\pi\to\chi\bar\chi$, and used them to solve the Boltzmann equation for the freeze-in yield.

The same interaction that controls the relic abundance also induces rare kaon decays with missing energy.
This connects the freeze-in target to rare-kaon searches at NA62 and KOTO.
Lower reheating temperatures require larger couplings to compensate for the thermal suppression of kaons in the bath, thereby shifting the freeze-in target toward the sensitivity range of current and future rare kaon experiments.

The differential missing-mass spectrum provides an additional handle on the DM interpretation.
The size and shape of the excess in $K_L\to\pi^0+\mathrm{invisible}$ are correlated with both $m_\chi$ and $T_{\rm RH}$.
Future high-statistics measurements of rare kaon decays could help distinguish a kaon-induced DM signal from the SM background.
Kaon-driven freeze-in therefore provides a concrete example in which the cosmological origin of light, feebly interacting DM can be tested through flavor experiments.

In this work, we have restricted our analysis to the vector-current operator as a simple benchmark scenario. 
Other dimension-six operators can change the correlation between the freeze-in abundance and rare kaon decay searches. 
For instance, scalar operators such as $(\bar{s}d)(\bar{\chi}\chi)$ are governed by the $K\to\pi$ scalar form factor, and can therefore lead to different decay spectra and freeze-in efficiencies. 
We have also assumed instantaneous reheating, so that the DM production begins at $T_{\rm RH}$. 
In non-instantaneous reheating scenarios, DM is also produced before radiation domination~\cite{Co:2015pka, Drees:2017iod, Banerjee:2022fiw}, which reduces the coupling required to reproduce the observed relic abundance.
A detailed investigation of these possibilities is left for future work.

\section*{Acknowledgement}
This work is supported by JSPS KAKENHI Grant Numbers 22K21347 [M.E.].
The work of T.Y. is supported in part by JST SPRING, Grant Number JPMJSP2138. 
T.Y. thanks KEK Theory Center for the hospitality during the visit, where part of this work was carried out.
\bibliography{ref}
\bibliographystyle{JHEP}

\end{document}